\documentclass{article}
\usepackage{frascatiphys,here,graphicx,subfigure}
\begin{document}
\title{ CHIRAL EXTRAPOLATION OF LIGHT RESONANCES \\ FROM UNITARIZED
CHIRAL PERTURBATION THEORY
}
\author{
J. R. Pel\'aez$^1$, C. Hanhart$^2$ and G. R\'{\i}os$^1$ \\
{$^1$\em Departamento de F\'{\i}sica Te\'orica II. Universidad Complutense}\\
{\em 28040 Madrid. Spain.} 
\\
{$^2$\em Instit\"ut f\"ur Kernphysik (Theorie), Forschungzentrum J\"ulich,} \\
{\em D-52425 J\"ulich, Germany}
}
\maketitle
\baselineskip=11.6pt
\begin{abstract}
  Both scalar and vector light resonances can be generated from the
   unitarization of one-loop chiral perturbation theory. This amounts
   to using in a dispersion relation  the chiral expansion,
   which incorporates the correct QCD quark mass dependence.  We can
   thus predict the quark mass dependence of the poles associated to
   those light resonances. Our results compare well with some recent
   lattice results for the $\rho(770)$ mass and can be used as a
   benchmark for future lattice results on the $\rho(770)$ or the
   $f_0(600)$ also known as the $\sigma$.
\end{abstract}
\baselineskip=14pt

\section{Introduction}

Light hadron spectroscopy at low energies lies beyond the realm of
perturbative QCD, although lattice QCD provides, in principle, a
rigorous way to extract non--perturbative quantities from QCD.
However, current lattice results are typically still done for
relatively high quark masses. Thus, in order to make contact with
experiment, appropriate extrapolation formulas need to be derived. This
is typically done by using Chiral Perturbation Theory
(ChPT)\cite{chpt1}, which provides a model independent description of
the dynamics of the lightest mesons, namely, the pions, kaons and
etas, which are identified with the Goldstone Bosons (GB) associated
to the QCD spontaneous Chiral Symmetry. Hence, ChPT is built out of
only those fields, as a low energy expansion of a Lagrangian whose
terms respect all QCD symmetries, and in particular its symmetry
breaking pattern. Actually, this chiral expansion becomes a
series in momenta and meson masses, generically $O(p^2/\Lambda^2)$,
when taking into account systematically the small quark masses of the
three lightest flavors that can be treated perturbatively.  
The chiral expansion scale is $\Lambda\equiv 4 \pi f_\pi$, where
$f_\pi$ denotes the pion decay constant.
  ChPT is renormalized order by order by absorbing loop
divergences in the renormalization of parameters of higher order
counterterms, known as low energy constants (LEC).
Their  values depend
on the specific QCD dynamics, and have to be determined either from
experiment or from lattice QCD --- they cannot be calculated from perturbative QCD.

The relevant remark for us is that, thanks to the fact that ChPT has
the same symmetries than QCD and that it couples to different kind of
currents in the same way, the orthodox ChPT expansion
 provides a {\it systematic and model
  independent} description of how the observables depend on some QCD
parameters.  This is the case for the leading dependence on the number
of colors $N_c$  and, more important for our purposes here, the
dependence on the quark masses, which can be implemented systematically up to the desired order in the orthodox ChPT expansion.

In this work we focus on the two lightest resonances
of QCD, the $\rho (770)$ and the $f_0(600)$.
It is therefore enough to work with the two lightest quark
flavors $u,d$ in the isospin limit of an equal mass that we take as
$\hat m=(m_u+m_d)/2$. The pion mass is given by an
expansion $m_\pi^2\sim \hat m+...$ (see \cite{chpt1} for details).
Therefore, studying the quark mass dependence is equivalent to study
the pion mass dependence. In $\pi\pi$ scattering at NLO within SU(2)
ChPT only four LECs $l_1,\cdots,l_4$ appear.  Of course, when changing
pion masses we have to take into account that amplitudes are
customarily written \cite{chpt1} in terms of the $\mu$ independent
LECs $\bar l$ \cite{chpt1} and the physical pion decay constant
$f_\pi=f_0\left(1+\frac{m_\pi^2}{16\pi^2 f_0^2}\,\bar
  l_4+\cdots\right)$ that depend explicitly on the pion mass, $m_\pi$.

\section{Unitarization and dispersion theory}

$S$ matrix unitarity implies, for physical values
of $s$, that elastic $\pi\pi$ scattering partial waves $t(s)$
of definite isospin $I$ and angular momenta $J$ should 
satisfy 
\begin{equation}
{\rm Im }\, t(s)=\sigma \vert t(s)\vert^2 \Rightarrow 
{\rm Im }\,\frac{1}{ t(s)}=-\sigma(s), \qquad {\rm with} \quad
\sigma(s)=2 p/\sqrt{s}
\label{unit}
\end{equation}
 and $p$ is the CM momenta.
Thus $\vert t^{IJ}\vert\leq1/\sigma$, and
interactions are said to become strong precisely when this unitarity bound is saturated. 

However, the ChPT
low energy expansion $t\simeq t_2+t_4+...$,
where $t_{2k}\equiv O(p/(4\pi f_\pi))^{2k}$,
can only satisfy unitarity perturbatively, i.e:
\begin{equation}
{\rm Im }\, t_2=0, \qquad
{\rm Im }\, t_4=\sigma t_2^2, \qquad {\rm etc...}
\end{equation}

The one-channel Inverse Amplitude Method (IAM)
\cite{Truong:1988zp,Dobado:1996ps} is a unitarization technique that 
can be derived within a ``naive'', intuitive, approach
by noting that eq.(\ref{unit}),
 {\it fixes the imaginary part of the inverse amplitude exactly}.
If we then use ChPT to write ${\rm Re}\, t^{-1}\simeq t_2^{-2}(t_2+{\rm Re}\, t_4+ ...)$, we find 
\begin{equation}
t\simeq=\frac{1}{{\rm Re}\, t^{-1}-i \sigma}=\frac{t_2}{1- t_4/t_2}.
\label{IAM}
\end{equation}
However, the above derivation is just formal, since
the ChPT series can only be used at low energies.
The correct derivation uses dispersion theory,
and the fact that the ChPT series of $t$ and $1/t$ beyond
leading order have an analytic structure
with a ``physical cut'' from threshold to $\infty$
and a ``left cut'' from $-\infty$ to $s=0$. This leads to
the following dispersion 
relation \cite{Dobado:1996ps} for $t_4$ 
\begin{eqnarray}
t_4 = b_0+b_1s+b_2s^2+     
\frac{s^3}\pi\int_{s_{th}}^{\infty}\frac{\rm Im\,
t_4(s')ds'}{s'^3(s'-s-i\epsilon)}+LC(t_4),
\label{disp1}
\end{eqnarray}
where ``LC'' stands for a similar integral over the left cut and 
we have three subtractions to ensure convergence. A similar dispersion relation
can be written for the function $G\equiv t_2^2/t$, by simply replacing $t_4$ by $G$
and changing the name of the subtraction constants.
Since $t_2$ is real, the  functions $G$ and $t_4/t_2^2$
have {\it exactly} opposite integrals over the physical cut.
Their subtractions constants are the value of these functions
at $s=0$ where 
the ChPT expansion is safe.
And finally, they also have opposite left cut contributions
up to NNLO ChPT. 
Such an approximation on the left cut is, of course, only justified for small
$\vert s\vert$, 
but due to the three subtractions this is precisely the 
region that dominates the left cut integrals.
Therefore, the IAM derivation is exact for the integrals over
 the elastic region 
and uses ChPT only where it is well justified. The IAM
is even more justified if used sufficiently far from the left cut,
as it is usually done, due to the additional $1/(s-s')$ suppression. 

In the scalar channels there are also contributions to the dispersion relation 
coming from poles in $1/t$ due to the so-called Adler zeros
located well below threshold. Such contributions lead, formally, to $O(p^6)$
corrections in the IAM,
and are customarily neglected, leading to the standard IAM we have justified
above. However, if not taken into account, the Adler zeros do not appear in 
the correct place and also unphysical poles occur in the IAM.
Still, the influence of these unphysical 
poles is very localized around those Adler zeros and the standard IAM
can be used safely for energies sufficiently far from the Adler zeros. 

Nevertheless, in the next section
we will show that resonance poles move into the subthreshold 
region for sufficiently large pion masses and it is thus
relevant to include the pole contributions
and use a slightly modified IAM, whose results 
agree with those of the standard IAM, except in the 
subthreshold region, where the modified version is more reliable. 
Such modified IAM has already been 
built and
used in \cite{FernandezFraile:2007fv}, 
by adding an {\it ad hoc} $O(p^6)$ piece to the ${\rm Re}\, t^{-1}$
expansion, 
 within the ``naive derivation'' explained just before Eq.(\ref{IAM}).
A rigorous dispersive derivation 
is in preparation \cite{modIAM}.

In summary, there are no model dependences in the approach, 
but just approximations to a given order in ChPT.
Remarkably, the simple formula of the elastic IAM, Eq.(\ref{IAM}), (or the slightly modified one
to work in the subthreshold region),
 while reproducing the ChPT expansion at low energies,
 is also able to generate both the  $\rho(770)$ and $f_0(600)$ 
resonances with values of the LECS compatible with 
standard ChPT \cite{Pelaez:2004xp,Guerrero:1998ei}.
In other words, the IAM generates the poles \cite{Dobado:1996ps,Pelaez:2004xp}
associated to these resonances in the second Riemann
sheet. The fact that resonances are {\it not introduced by hand} but generated
from first principles and data, is relevant because the 
existence and nature of scalar resonances is the
subject of a long-lasting intense debate.

To be precise, the IAM, when reexpanded, reproduces the orthodox
ChPT series up to the order to which the input amplitude was evaluated
and, in particular the quark mass dependence agrees with that of ChPT 
up to that order.
A few  of the higher order terms are produced correctly
by the unitarization but not the complete series--- for
a discussion of this issue for the scalar pion formfactor see 
Ref.\cite{Gasser:1990bv}. However, the formalism just described still provides
us with a fair estimate of the quark mass dependence of the resonance
properties. 
In this case, we can study, 
 without any a priory assumption, the dependence 
  of the resonances positions on QCD parameters like the number
of colors $N_c$ \cite{Pelaez:2006nj} or  their dependence on the quark masses 
up to a given order in ChPT. 
\section{Results}

As commented above, since, for the moment, we are only interested
in resonances appearing in elastic $\pi\pi$ scattering, 
we can unitarize  $SU(2)$ ChPT at one-loop. 
For the LECS we take 
$l_3^r=0.8\pm 3.8,\,l_4^r=6.2\pm 5.7$, directly from \cite{chpt1},
(since the partial waves where the $\sigma$ and $\rho$ appear are not
very sensitive to these two constants), whereas
we use $l_1^r=-3.7\pm 0.2,\,l_2^r=5.0\pm 0.4$ obtained from an IAM fit
to data up to the resonance region. All LECS are evaluated at $\mu=770\,{\rm MeV}$,
and are in fairly good agreement with standard values.

The highest value of $m_\pi$ we can use is limited since we
do not want to spoil the chiral expansion and we want to have some elastic 
$\pi\pi$ regime below the $K \bar K$ threshold. 
A mass of $m_\pi \leq500\,{\rm MeV}$ satisfies both criteria
since we know SU(3) ChPT still works fairly well with a kaon mass that high,
and also because if we increase the pion mass to 500 MeV, the kaon mass becomes
$\simeq$ 600 MeV, and $\pi\pi$ scattering is still elastic for 200 MeV,
before reaching
the two-kaon threshold. To reach higher masses we would need a coupled-channel IAM,
which is feasible, but lies beyond our present scope.

\begin{figure}
     \begin{center}
         {\includegraphics[scale=0.27,angle=-90]{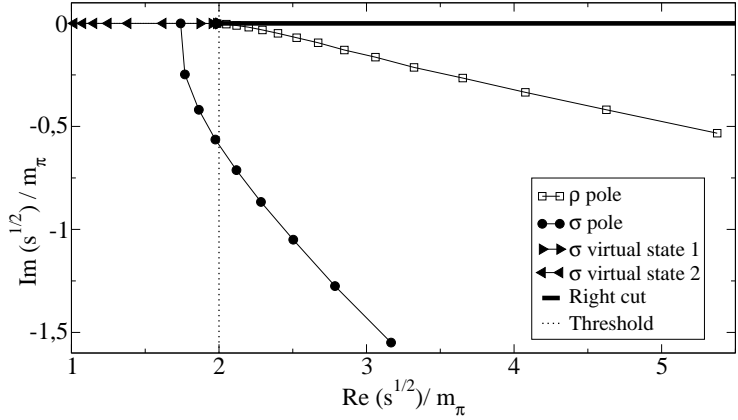}}
\caption{$\rho$ and $\sigma$ complex plane pole 
movement with increasing pion mass.
  To ease the comparison of the pole position 
relative to the two-pion threshold we normalize
   by the pion mass that is changing. Note how the sigma pole moves 
  toward the real axis below threshold where it splits in two virtual states,
  whereas the $\rho$ pole just moves toward threshold.}
 \label{Fig:F1}
     \end{center}
 \end{figure}

Thus, in Fig.\ref{Fig:F1} we show the $\rho $ and $f_0(600)$ poles
movement in the second Riemman sheet as $m_\pi$ increases.  Note that
in order to follow easily the pole movement relative to the two-pion
threshold, which is also increasing, we express all quantities in
units of $M_\pi$, so that the two-pion threshold is shown fixed at
$\sqrt{s}=2$.  In this way we clearly see that both poles move closer
to the two-pion threshold.  Let us recall that for narrow resonances,
their mass $M$ and width $\Gamma$ are related to the pole position as
$\sqrt{s_{pole}}\simeq M -i\Gamma/2$ and customarily this notation is
also kept for broader resonances.  Hence, both the $\sigma$ and $\rho$
widths decrease for increasing $m_\pi$, partly due to phase space
reduction.  In particular, the $\rho$ pole moves toward the real axis
and just when the threshold is reached it jumps into the real axis on the
first sheet, thus becoming a traditional bound state, while its
conjugate partner remains on the second sheet practically at the very
same position as the one in the first.  In contrast, when the $\sigma$
mass reaches the two-pion threshold, its poles remain on the 
second sheet with a non-zero imaginary part before 
they meet on the real axis and become virtual states. 
As $m_\pi$ increases
further, one of those virtual states moves towards the threshold and
jumps onto the first sheet, whereas the other one remains in the
second sheet.  Although, of course, this happens for very large
values of $m_\pi$, such an analytic structure, with two very asymmetric
poles in different sheets of an angular momentum zero partial wave,
is a signal for a prominent molecular component
\cite{Weinberg}. 
Differences between P-wave and S-wave pole movements were also found
within quark models \cite{vanBeveren:2002gy}, the latter showing also two 
second sheet poles on on the real axis below threshold.

 In Fig.2 (left) we show in detail, the growth of the $\sigma$ and
 $\rho$ masses, starting from the chiral limit up to $\sim$ 500 MeV.
 We find that both the $\rho$ and $\sigma$ mass increase, but that of
 the $\sigma$ grows faster, until it splits in two virtual states.
 Then one mass keeps growing, whereas the other one decreases.

Finally in Fig.2 (right) we show our central value result 
for the $\rho$ mass dependence on $m_\pi$ compared 
with some recent lattice results \cite{Aoki:1999ff}.
Despite our results refer to the $\rho$ ``pole-mass'' definition and that those results
on the lattice have a zero width for the $\rho$, we see a reasonable qualitative agreement
between both results, although our dependence seems to be somewhat steeper.
We have some preliminary indications that if we decrease the $\rho$ width in our
approach (by taking the large $N_c$ limit of ChPT), 
we reproduce even better those lattice results.

 \begin{figure}
     \begin{center}
         {\includegraphics[scale=0.21,angle=-90]{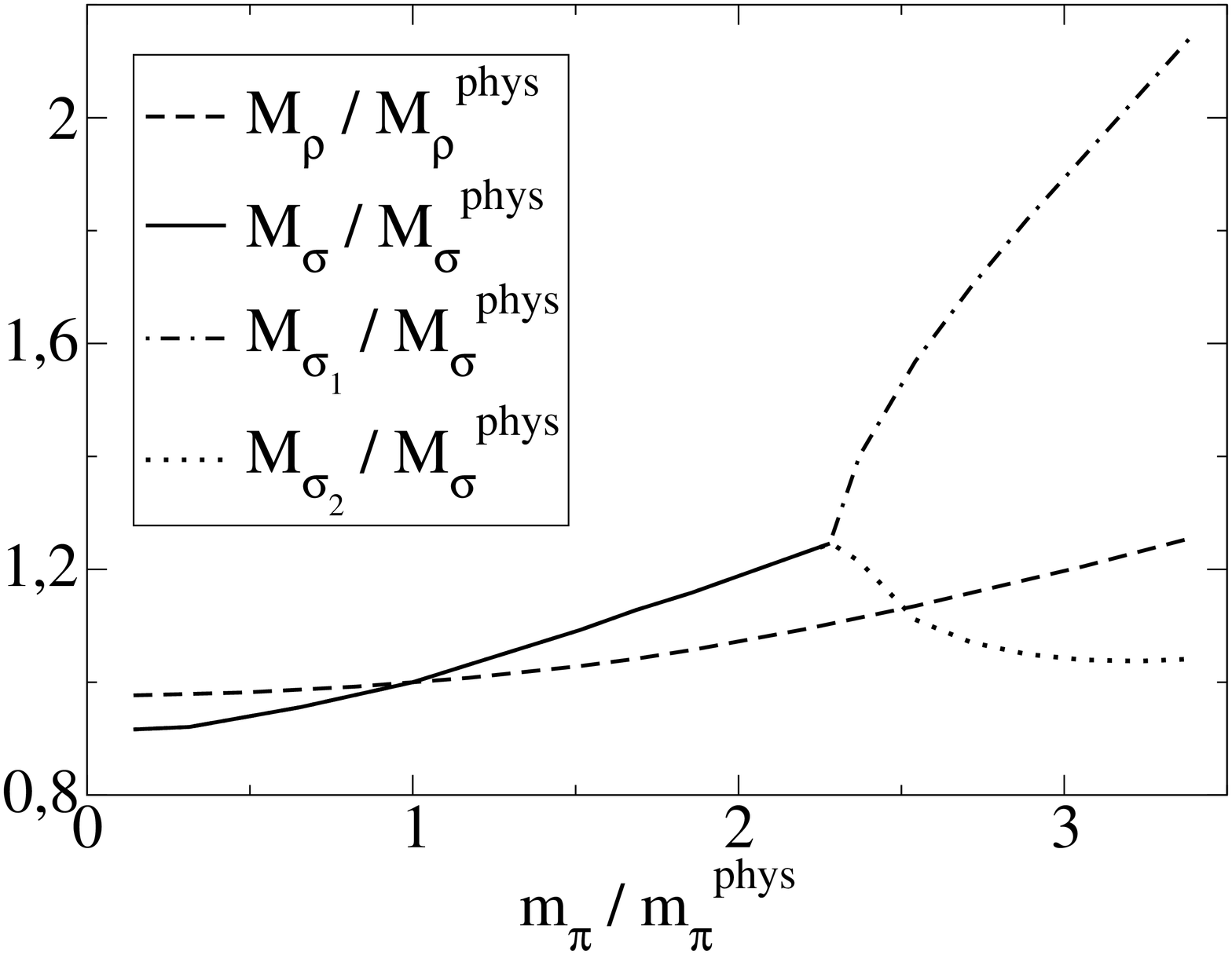}
         \includegraphics[scale=0.21,angle=-90]{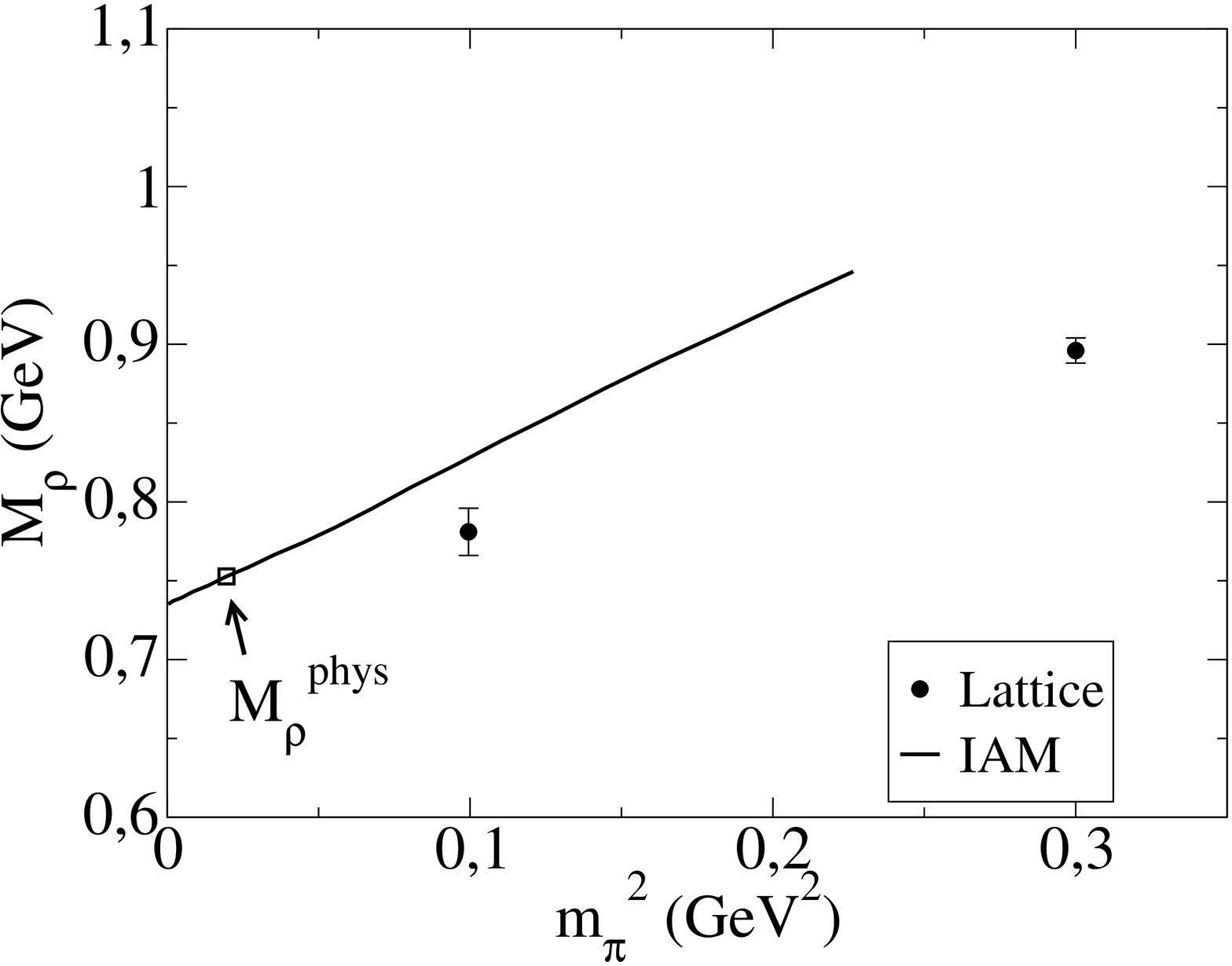}}
\caption{left:$\rho$ and $\sigma$ mass dependence on $m_\pi$. Note that the sigma pole
splits in two virtual poles, denoted $\sigma_1$ and $\sigma_2$ for sufficiently high $m_\pi$.
Right: IAM $\rho$ mass dependence on $m_\pi$ versus
some recent lattice results (see text).}
 \label{Fig:F2}
     \end{center}
 \end{figure}

 A publication with further details
 is in preparation \cite{inprep2}
 including results of the $f_0(600)$ and $\rho(770)$ mass and 
width evolution with the pion mass
 as well as a comparison with other works and lattice results. 
Estimates of uncertainties and possibly 
 an extension to the SU(3) coupled channel case are in progress.

\section{Acknowledgements}
We thank U. G-. Mei{\ss}ner and E. Van Beveren for
their suggestions. 
J.R.P. and G.R. research partially funded by Banco Santander/Complutense
contract PR27/05-13955-BSCH and Spanish CICYT contracts
FPA2007-29115-E, FPA2005-02327, BFM2003-00856.
J.R.P. and C. Hanhart research is part of the EU integrated
infrastructure initiative HADRONPHYSICS PROJECT,
under contract RII3-CT-2004-506078.

\end{document}